\documentclass [11pt]{article}
\usepackage[dvips]{color}
\usepackage{epsfig}
\usepackage[normalem]{ulem}

\def\beq{\begin{equation}}
\def\eeq{\end{equation}}
\def\beqa{\begin{eqnarray}}
\def\eeqa{\end{eqnarray}}
\def\d{{\rm d}}
\def\ttimes{{\scriptstyle \times}}

\def\rU{{\mbox{\tiny \rm U}}}

\def\rint{{\mbox{\tiny \rm int}}}

\begin{document}
\baselineskip0.6cm plus 1pt minus 1pt
\tolerance=1500

\begin{center}
{\LARGE\bf Thermodynamics in rotating systems  --- analysis of selected examples}
\vskip0.4cm
{ J. G\"u\'emez$^{a,}$\footnote{guemezj@unican.es},
M. Fiolhais$^{b,}$\footnote{tmanuel@teor.fis.uc.pt}
}
\vskip0.1cm
{\it $^a$ Departamento de F\'{\i}sica Aplicada}\\ {\it Universidad de
Cantabria} \\ {\it E-39005 Santander, Spain} \\
\vskip0.1cm
{\it $^b$ Departamento de F\'\i sica and Centro de
F\'\i sica Computacional}
\\ {\it Universidade de Coimbra}
\\ {\it P-3004-516 Coimbra, Portugal}
\end{center}

\begin{abstract}
We solve a set of selected exercises on rotational motion requiring a mechanical and thermodynamical analysis.
When non-conservative  forces or thermal effects are present, a complete study must use the first law of thermodynamics together with the
Newton's second law. The latter is here better expressed in terms of
an  `angular' impulse-momentum equation (Poinsot-Euler equation), or, equivalently, in terms of a  `rotational' pseudo-work-energy equation.
Thermodynamical aspects in rotational systems, when e.g. frictional forces are present or when there is a variation of the rotational kinetic energy due to internal sources of energy,
are discussed.
\end{abstract}

\section{Introduction}
\label{sec:intro}

Rotations are present everywhere. Their mechanical treatment relies on the Newton's second law, whose formulation for rotations leads to the so-called Poinsot-Euler equation \cite{taylor59}.
In spite of the analogy that can be established between translations and rotations, some peculiarities associated with the rotational motion makes it sometimes more challenging for students than the translational motion.
For instance, the direction of the static friction force, which can be either the direction of the centre-of-mass motion of a rolling object or the opposite direction,
is usually presented as an example of such difficulties \cite{pinto01}. In addition, if there are kinetic frictional forces, or internal sources of energy
responsible, e.g. for the production of kinetic rotational energy, another type of problems arise,  namely on the   use of the energy balance equation
 (first law of thermodynamics) in addition to the Newton's second law.
Not so surprisingly, and in spite of their presence in everyday life and of their scientific interest,
examples of   rotational processes  with  mechanical energy dissipation (e.g. a sliding disc on an incline \cite{knudsen00}) or with production  of mechanical energy
(e.g. a fireworks wheel) are  scarce in the literature.

In a previous publication
\cite{guemez13} we presented a number of examples of systems in translation, ranging from mechanics to thermodynamics, to illustrate the applicability of the
(pseudo)work--energy equation  \cite{sherwood83} (actually, equivalent to Newton's second law \cite{penchina78}) versus the first law of thermodynamics \cite{sousa97}.
In this paper, in the same spirit,  we present  examples of rotational motion whose description also requires both mechanics and thermodynamics.

The classical dynamics of  bodies in translation is described using
Newton's second law which can be  expressed by \cite{guemez13,penchina78}
\beq
{\vec F}_{\rm ext}\  \d t = M \d {\vec v}_{\rm cm}\, , \ \ \ \  \mbox{or} \ \ \  {\vec F}_{\rm ext} \cdot \d {\vec r}_{\rm cm} = {1\over 2} M \d v_{\rm cm}^2  \ \label{eq-3}
\eeq
where ${\vec F}_{\rm ext}$ is the resultant of the external forces and ${\vec v}_{\rm cm}$ the center-of-mass velocity of the system of constant mass $M$.
The first expression in (\ref{eq-3}) is the impulse-momentum equation and, the second one, the centre-of-mass (pseudo-work) equation
 \cite{mallin92}.

For the rotation of a rigid body around a principal inertia axis containing the centre-of-mass, the   Newton's second law is better expressed by (Poinsot-Euler equation) \cite{taylor59,tipler04,varios}
\beq
\vec {\Gamma}_{\rm ext}\d t = I \d \vec{\omega}\, ,
\label{eq:in002a}
\eeq
where $\vec {\Gamma}_{\rm \, ext} = \sum_j \vec \tau_j^{\rm \, ext} $ is the resultant external torque,  $\vec \omega$ is the angular velocity around the axis and $I$ the constant moment of inertia with respect to that principal axis. This is an angular impulse-momentum equation whose evident analogy with the first equation (\ref{eq-3}) is rather appealing and always worthwhile  to point out in a classroom context.
However, for a constant $I$,
the previous equation is only valid when the internal forces produce a vanishing torque, which is the most common situation.  Otherwise equation (\ref{eq:in002a}) should include, on the left-hand side, the torque of the internal forces  which does not vanish if the two forces of an action-reaction pair are not along the same line but along two parallel lines.  
After multiplying both sides of equation~(\ref{eq:in002a}) by $\vec\omega$, and since the rotation is along a principal inertia axis ($\vec {\Gamma}_{\rm ext}$ and $\vec    \omega$ are, therefore, co-linear),
one obtains the equivalent to the second equation (\ref{eq-3}) for the restricted class of rotations considered here, namely
\beq
\Gamma_{\rm ext} \d \theta = {1\over 2}I \d \omega^2\,
\label{eq:in002b}
\eeq
where $\d \theta = \omega \d t$ is the infinitesimal angular displacement of the body.
Similarly to the second expression in equation  (\ref{eq-3}), one should note that
equation~(\ref{eq:in002b})
is not, in general, an expression of the `work-energy theorem' for rotations \cite{sherwood84,walker11c}.
Actually,  in general, $\Gamma_{\rm ext} \d \theta$  is  not   always a work \cite{young12}.
The association of $ \Gamma_{\rm ext} \d \theta$  always with real work is as erroneous as the association of the pseudo-work $ {\vec F}_{\rm ext} \cdot \d {\vec r}_{\rm cm}$, in equation~(\ref{eq-3}), always with real work.
Some  examples presented in this paper will emphasize this point. It should remain clear that equations (\ref{eq-3}) are always valid for translations of a constant mass body,
whereas equations (\ref{eq:in002a}) and (\ref{eq:in002b}) are valid for rotations under the specific circumstances mentioned above.

When thermal effects are present, they cannot be accounted for by any of the previous equations, either in the case of translations or rotations.  In such a case
one has to consider, additionally, the first law of thermodynamics  \cite{sousa97,erlichson84,serway02a}.
For the sake of completeness here we very briefly repeat  part of the discussion presented in \cite{guemez13}.
The internal energy infinitesimal
variation of any system, $\d U$, may receive contributions: i) from the variation of the internal kinetic energy, $\d K_{\rint}$ (it includes rotational kinetic energy and translational kinetic energy with respect to the centre-of-mass);
ii) from any internal work, $\d w_{\rint} = -\d \Phi$ ({\em i.e.} work possibly done by internal forces) \cite{mallin92}; iii)
from the internal energy variations related to temperature variations, expressed as $ M\, c\, \d T$ ($c$ is the specific heat); iv)  from the internal energy variations
related to chemical reactions \cite{atkins10}; v) from any other possible way not explicitly mentioned before.

It is well known that both work and heat contribute to the internal energy variation of a system.
For a general process on a macroscopic system, whose analysis needs to combine mechanics and thermodynamics \cite{jewett08v},
besides the centre-of-mass equation, such as the second expression in (\ref{eq-3}), or  (\ref{eq:in002b}), one should also consider the  equation \cite{guemez13}
\beq
\d {K}_{\rm cm} + \d { U} = \sum_j { {\vec F}_{{\rm ext}, j}}\cdot \ \d {\vec r}_j  \ + \ \delta { Q}
\label{totale}
\eeq
which is nothing but an expression of the  first law of thermodynamics \cite{erlichson84,besson01}:
this equation is complementary to the centre-of-mass equations (\ref{eq-3}) or (\ref{eq:in002a})-(\ref{eq:in002b}), but all are valid.
(The infinitesimal heat in (\ref{totale}) is denoted by $\delta Q $ because it is not an exact differential \cite{landsberg83}.)
Each term in the sum on the right-hand side of equation~(\ref{totale})
is work associated with each {\em external} force ${\vec F}_{{\rm ext}, j}$, and  $\d {\vec r}_j$
is the infinitesimal displacement of the  force ${\vec F}_{{\rm ext}, j}$ itself (not the displacement of the centre-of-mass). Therefore, $\delta  {W}_j= {\vec F}_{{\rm ext}, j} \cdot \d {\vec r}_j$ is always real work and not pseudo-work.  Note that an external force may have an associated pseudo-work (when there is displacement of the system centre-of-mass) without doing any work, if its application point does not move --- a common example is the force acting on the foot of a walking person \cite{guemez13c}.

We stress that, in expressing the first law by equation (\ref{totale}), one assumes that any translational kinetic energy with respect to the center-of-mass and the rotational energy should be included in the internal energy of the system, a point that will be illustrated with the examples of next sections.
Moreover, the range of applicability of equation~(\ref{totale}) is larger than the most known expression for the first law, $\d U=\delta W+\delta Q$, since it  can also be applied to thermal engines that move by themselves (a steam locomotive, a car, a person, etc.).

If dissipative forces are present, such as kinetic friction or dragging forces, it is not unambiguous  what their displacements are \cite{besson01,serway}. Therefore, we may have a problem in computing the work from its basic definition \cite{serway}. However, since dissipative work is equivalent to heat, we may always assume that the energy transfer associated with such forces is accounted by the last term in (\ref{totale}):  operationally we simply consider that any dissipative work, $\delta W_{\rm D}$, is thermodynamically equivalent  to a heat transfer $\delta Q=\delta W_{\rm D}$. We take this point of view and, as we shall see in various examples, this protocol does not affect the mechanical treatment (for which only pseudo-works matter) and even allows for a very clean analysis from both the first and the second laws of thermodynamics. The important point is to consider the energy transfer associated with dissipation and include it on the right-hand side of equation (\ref{totale}).

Both the second equation (\ref{eq-3}) and equation (\ref{totale}) refer to energy balances in processes but they express two quite different fundamental physical laws. The former, the so-called centre-of-mass energy equation,
is an alternative  way of stating Newton's second law, whereas the latter expresses the first law of thermodynamics  in its most general form. They are both simultaneously general and always valid,
therefore each one provides new information with respect to the other. Of course, if the problem is out of the scope of thermodynamics, the two equations
are then equivalent or, in other words, they become the same equation. This happens when there is no internal energy variation, no heat transfers and when the displacement
of the forces equals the displacement of the centre-of-mass. The sum on the right-hand side of equation~(\ref{totale}),
$   \sum_j { {\vec F}_{{\rm ext}, j}}\cdot \ \d {\vec r}_j$, then simply becomes  $\vec F_{\rm ext}\cdot \ \d {\vec r}_{\rm cm}$, where $\vec F_{\rm ext}= \sum_j { {\vec F}_{{\rm ext}, j}}$ is the resultant of the external forces, hence (\ref{totale}) becomes equal to the second equation (\ref{eq-3}).

Summarizing, within their limits of applicability, all above
equations, given in infinitesimal form, are  valid and, therefore, they should be  compatible. Equation (\ref{totale}) may provide the same information as the other energy equations if the problem under consideration is a purely mechanical one. However, if the situation lies in the scope of thermodynamics, that equation, which embodies the first law, provides new information with respect to the center-of-mass equation in (\ref{eq-3}) or with respect to equation (\ref{eq:in002b}). Of course, for a complete study, one
still needs the second law of thermodynamics \cite{leff12b}, which states that only processes compatible with a non-decrease of the entropy of the universe, $\Delta S_\rU \ge 0$, are allowed.

To illustrate the usage of the angular impulse-momentum equation (\ref{eq:in002a}) together with (\ref{eq:in002b}), or with the more general energy equation (\ref{totale}), we discuss, in the next sections, a set of examples combining mechanics and thermodynamics.
We selected a series of representative examples, ranging from mechanics to thermodynamics as in \cite{guemez13}, which illustrates the additional information that can be extracted from equation (\ref{totale}), with respect to Newton's equation. It is important to note that in the formulation of Newton's second law in terms of energy --- second equation in (\ref{eq-3}) and equation (\ref{eq:in002b}) --- in general it is the pseudo-work that matters, whereas in the first law of thermodynamics it is the work that matters. This is a clear difference between (\ref{totale}) and the preceding equations, which is very natural since they correspond to two distinct physical laws. This subtle difference is probably not always adequately mastered.

This paper, following the structure used in \cite{guemez13}, is organized as follows.
In  section~\ref{sec:falling} we study a descending pulley, connected to a rotating axially fixed one, the whole system being a mechanical energy conserving one.  In this case,  the system is a pure mechanical one and equation (\ref{totale}) does not apport additional information with respect to the mechanical equations.
In section~\ref{sec:rotatingdisc} it is analyzed the motion of a disc acted upon by a pair of opposite forces and also subjected to friction forces (with a consequent mechanical energy dissipation).  We treat the dissipative work as heat when we apply equation (\ref{totale}), as mentioned above. In section \ref{sec:fire} we consider a simplified fireworks wheel as an example of a system where there is production of  rotational kinetic energy.
In Section \ref{sec:joule} we discuss, in the mechanical-thermodynamical perspective, the historical wheel paddle Joule's apparatus  to illustrate the `mechanical equivalent of heat'  In the final section we present the conclusions.

\section{Falling and rotating pulley}
\label{sec:falling}

\begin{figure}[htb]
\begin{center}
\hspace*{-0.5cm}
\includegraphics[width=5.5cm]{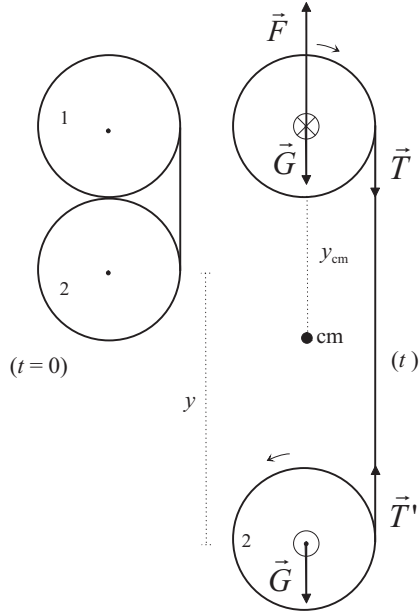}
\end{center}
\vspace*{-0.5cm}
\caption[]{\label{fig:rotatingdiscdescendingdisc} \small Falling rotating pulley attached to an axially fixed rotating pulley, at $t=0$ and at a generic $t$.
 Indicated are the tension forces on the pulleys transmitted by the rope, $\vec T $ and $\vec T'$, the weights, $\vec G$, and the force, $\vec F$, exerted by the support on the top pulley.
} 
\end{figure}

We first consider  a translational-rotational motion where there is conservation of mechanical energy.
In figure~\ref{fig:rotatingdiscdescendingdisc} the system and the process are sketched: a pulley of mass $M$ and radius $R$,  rotates and falls down attached by a rope to a similar axially fixed rotating pulley  (\cite{knudsen00} p.~307).
The rotation of the upper pulley or disc (disc 1) is frictionless and the moment of inertia for the rotation around the axis is assumed as $I={1\over 2}MR^2$.

 We first consider two separate systems: pulley 1 and pulley 2.
The forces are indicated in the figure --- the two tensions are constant and equal in magnitude, $T'=T$,  which establishes a connection between the two systems; the weight of each disc is  $G=Mg$ and, since disc 1 has no translational motion,  $F=Mg+T$. For the rotation of disc 1,  assuming that the discs are at rest for $t=0$, equations (\ref{eq:in002a}) and (\ref{eq:in002b}) lead, after a trivial integration, to
\beq
\left\{
\begin{array}{rl}
I \omega  &=  T R \,t  \vspace*{0.3cm}\\
\ \ {1\over 2} I \omega^2 &= T R \,\theta
\end{array}
\right.
\label{8670r}
\eeq
where  $\omega$ is the angular velocity at instant $t$ and $\theta$ is the corresponding angular displacement of the disc.
The angular acceleration is constant, $\alpha= TR/I$, and $\omega=\alpha t$. From equations (\ref{8670r})  one readily obtains $\omega=2 \theta / t$ and, therefore,
$\theta = {1 \over 2} \alpha t^2$. Because $T'=T$, rotation equations for disc 2 are the same as for 1, so all the previous results hold for disc 2 ($\theta$, $\omega$ and $\alpha$ are the same for both pulleys).

Regarding the translation of disc 2, equations (\ref{eq-3}) lead to
\beq
\left\{
\begin{array}{rl}
M v  &=  (Mg-T)\, t  \vspace*{0.3cm}\\
\ \ {1\over 2} M v^2 &= (Mg-T)\, y
\end{array}
\right.
\label{8670s}
\eeq
where  $v$ is the velocity of the center of disc 2 at instant $t$ and $y$ is the vertical displacement of that disc. The disc falls down with constant acceleration,
$a=(Mg-T)/M$, and $v=a t$. From equations (\ref{8670s})  one obtains $v=2y  / t$ and, therefore,
$y = {1 \over 2} \, a t^2$.

Assuming that the rope does not slide on the pulleys' rim, the connection between the translation and rotations is expressed by the geometrical relation $y=2R\theta$,
which yields $v=2R\omega$ and $a=2R\alpha$. Using these relations in (\ref{8670r}) and  (\ref{8670s}) one obtains
\beq
T={1\over 5}Mg, \ \ \ \ F={6 \over 5}Mg, \ \ \ \ a=  {4\over 5} g, \ \ \ \ v= \sqrt{{8\over 5} g y}\, .
\label{trgh}
\eeq

 Next we consider the translation of the system as a whole (pulley 1 + pulley 2). Equations (\ref{eq-3}) lead to
\beq
\left\{
\begin{array}{rl}
2 M v_{\rm cm}  &=  (2Mg-F)t  \, , \vspace*{0.3cm}\\
\ \ {1\over 2} 2 M v_{\rm cm}^2 &= (2Mg-F)y_{\rm cm}  \, .
\end{array}
\right.
\label{8670t}
\eeq
The acceleration of the center-of-mass,  $a_{\rm cm}= v_{\rm cm}/t$, is readily obtained as well as the center-of-mass velocity and the vertical displacement:
\beq
a_{\rm cm}= {2\over 5} g, \ \ \ \ v_{\rm cm}= {v\over 2},   \ \ \ \ \ \    y_{\rm cm}= {y\over 2}\, .
\label{tys}
\eeq

Let's now investigate what does equation (\ref{totale}), applied to the whole system, tell us in this process. The integrated form of that equation reads
\beq
\Delta K_{\rm cm} + \sum_k \Delta U_k = \sum_j W_j^{\rm ext} + Q
\label{totalei}
\eeq
The first term, $\Delta K_{\rm cm}$, is the variation of center-of-mass kinetic energy, whose calculation is straightforward using (\ref{tys}).  The rotational energy of the discs is part of the internal energy. We denote the associated energy variation by $\Delta U_{\rm R}$. On the other hand, the translational kinetic energy of the system with respect to its center-of-mass is also part of the internal energy of the system as a whole, and we denote that energy variation by
$\Delta U_{\rm K}$.  In the absence of dissipative processes or internal sources of energy there is no temperature change, hence the internal energy variation is just due to the variation of the internal kinetic energies, namely $\Delta U=\Delta K_{\rm int}=\Delta U_{\rm R}+\Delta U _{\rm K}$.  Regarding the right hand side of (\ref{totalei}) the work of the external forces is the work of the gravitational force on disc~2, $W_G$ (the tensions on the rope are internal forces and, moreover, their total work is zero). Altogether, and taking into account the previous results, one has
\beq
\begin{array}{rl}
\Delta K_{\rm cm}  &=  {1\over 2} 2M v_{\rm cm}^2 = {1\over 4} M v^2 \vspace*{0.3cm}\\
\Delta U_{\rm K} &= {1\over 2} M (0-v_{\rm cm} )^2 + {1\over 2} M (v-v_{\rm cm} )^2 = {1\over 4} M v^2 \vspace*{0.3cm}\\
\Delta U_{\rm R} & = 2 \ttimes {1\over 2} I \omega^2 =  {1\over 8} M v^2 \vspace*{0.3cm}\\
W_G & = Mgy\, .
\end{array}
\label{8670x}
\eeq
Inserting in equation (\ref{totalei}) one obtains
\beq
{5 \over 8 } M v^2 = Mgy + Q\, .
\label{aavcf}
\eeq
Using here the last expression in (\ref{trgh}) for $v$, one immediately concludes that $Q=0$, an expected result consistent with the assumption of the non sliding condition. Therefore, one also concludes that the entropy of the universe does not increase, the process is reversible and  the mechanical energy remains constant.  In this case equation (\ref{totale}) and the second equation (\ref{eq-3}) are equivalent.

 If there were  dissipative friction forces, for instance in the rotation of the axially fixed pulley or between the rope and the pulleys, the mechanical treatment would be more complicated (due to the presence of new forces and torques) and so would the thermodynamical treatment. In that case, $Q\not= 0$ corresponding to a heat transfer to the surroundings and a possible temperature increase of the pulleys. This will be the case in the next example where we study a rotating system in which kinetic friction forces are present.
Though equation (\ref{aavcf}) gets modified if there are dissipative forces, its present form allows for some qualitative analysis if we assume a heat such that $Q<0$ (heat dissipated in the surrounding). In   that case, $Mgy>{5 \over 8 } M v^2  $, i.e. for the same distance traveled by centre-of-mass the velocity would be lower than it actually is when no dissipative processes are present.

\section{Rotating disc with friction}

\label{sec:rotatingdisc}

\begin{figure}[htb]
\begin{center}
\hspace*{-0.5cm}
\includegraphics[width=14cm]{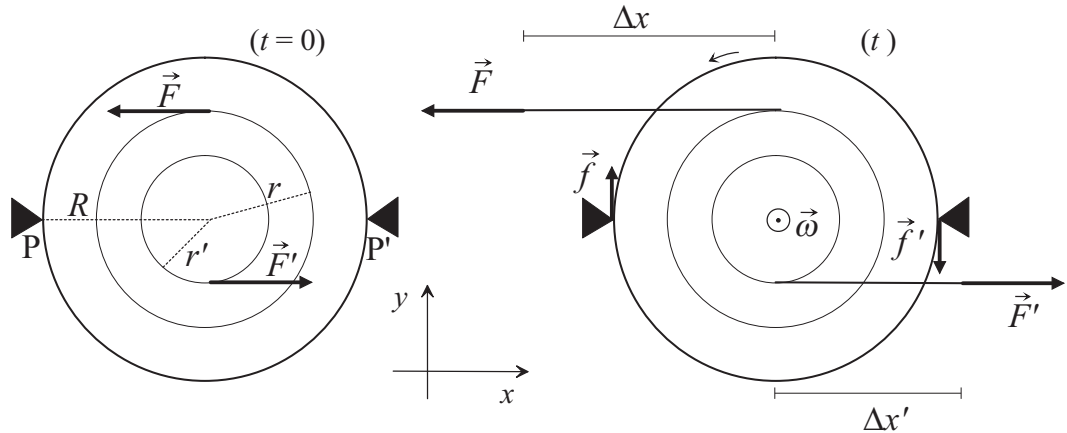}
\end{center}
\vspace*{-0.5cm}
\caption[]{\label{fig:rotatingdiscfriction} \small Rotating disc with friction, at $t=0$ and at a generic $t$.
 The vectors are the tension forces transmitted by the ropes, $\vec F$ and $\vec F'$, and the friction forces $\vec f$ and $\vec f'$.
} 
\end{figure}

An axially fixed disc of radius $R$, mass $M$ and moment of inertia $I$ may freely rotate  around its axis. It is acted upon by two horizontal forces, $\vec F$ and $\vec F '$, equal in magnitude and opposite, applied on (massless) ropes  wrapped around two narrow pulleys of radii $r$ and $r'$,
as shown in fig. \ref{fig:rotatingdiscfriction}.  On the other hand, two vertical friction forces, $\vec f$ and $\vec f '$, act on the rim of the disc at P and P$'$.
The cartesian components of the forces are: $\vec F=(-F,0,0)$ and $\vec F '=(F,0,0)$; $\vec f=(0,f,0)$ and $\vec f '=(0,-f,0)$ --- hence, there is no translational motion, i.e. $\d K_{\rm cm}=0$.

 The system is the disc and the massless ropes.
 Equations (\ref{eq:in002a}) and (\ref{eq:in002b}) are readily integrated (all forces are constant) leading to
\beq
\left\{
\begin{array}{rl}
I \omega  &=  \left[ -2fR + F (r+r')\right]  t  \, , \vspace*{0.3cm}\\
\ \ {1\over 2} I \omega^2 &= \left[ -2fR + F (r+r')\right]  \theta \, .
\end{array}
\right.
\label{86r}
\eeq
As in the previous example, $\omega =2 \theta/t$. Since $F$ and $f$ are constants, the angular acceleration is constant, $\alpha=\omega/t$, and given by
\beq
\alpha=\frac{F(r+r')-2fR}{I}
\eeq
For given $F$ and $f$ (which is such that $\alpha$ is non-negative) this acceleration can be determined. On the other hand, the angular velocity and the angular displacement at instant $t$
are $\omega=\alpha t$ and $\theta={1\over 2} \alpha t^2 $, respectively.

Next, let us consider the energy equation (\ref{totale}) for the system. There is no center-of-mass displacement and, hence, there is no translational energy with respect to the center-of-mass. Therefore,
the internal energy variation solely results from the rotational energy increase, ${1\over 2}I \omega^2 $, and from a possible temperature variation expressed by $Mc\Delta T$.
Equation (\ref{totale}) allows us to write
\beq
\label{eq:rdf003}
{1\over 2} I \omega^2  + M c \Delta T = F  (r +  r') \theta + Q\, ,
\eeq
where the work of the forces  transmitted by the ropes (whose displacements are  $\Delta x=r\theta$ and $\Delta x'=r'\theta$) is taking into account on the right-hand side.  The friction forces are always applied at the same points and it can be argued that they don't do any work \cite{halliday01} accountable by the first term on the right-hand side of (\ref{totale}), as mentioned in the Introduction. Instead, we take the point of view that this dissipative work should be taken into account  by the second term on the right-hand side of (\ref{totale}).
Using the second equation (\ref{86r}) in (\ref{eq:rdf003}) one concludes that $Q= -2fR\theta+ Mc \Delta T$. If we define $W_{\rm D}=2fR\theta>0$, we can still write $W_{\rm D}= Mc \Delta T - Q$ and the interpretation is clear: part of the energy resulting from friction, $W_{\rm D}$, is absorbed by the system which increases its temperature and, therefore, its internal energy; the other part is a heat transfer to the surrounding, which is suppose to be a heat reservoir. If the temperature variation is ignored hence, $Q=-W_{\rm D}=-2fR\theta$ is the heat exchanged with the heat reservoir that surrounds the disc. In that case, the increment of the entropy of the universe is
$\Delta S_{\rm U} = -{Q\over T} =   {2 f R \theta\over T} > 0$ (if the body's temperature  increase due to friction is considered, one certainly has a different entropy increase but one still has $\Delta S_\rU>0$).
We note that the  force-displacement product $(2f)(R\theta)$  was not considered as work strictly speaking~\cite{wolfson07}
but rather as thermal energy~\cite{besson03}.

This is a mechanical energy dissipating process and it is an irreversible one.
 If the friction forces  vanish, $f=0$, there would be no entropy increase and the process would be reversible. In this case the energy equation~(\ref{totale}) would not provide any additional information in comparison to (\ref{eq:in002b}).  Finally, we note that if the friction forces were in directions opposite to those represented in figure~\ref{fig:rotatingdiscfriction}, $f<0$, and the entropy of the universe would decrease, which is forbidden by the second law.

\section{Fireworks wheel}

\label{sec:fire}

\begin{figure}[htb]
\begin{center}
\hspace*{-0.5cm}
\includegraphics[width=14cm]{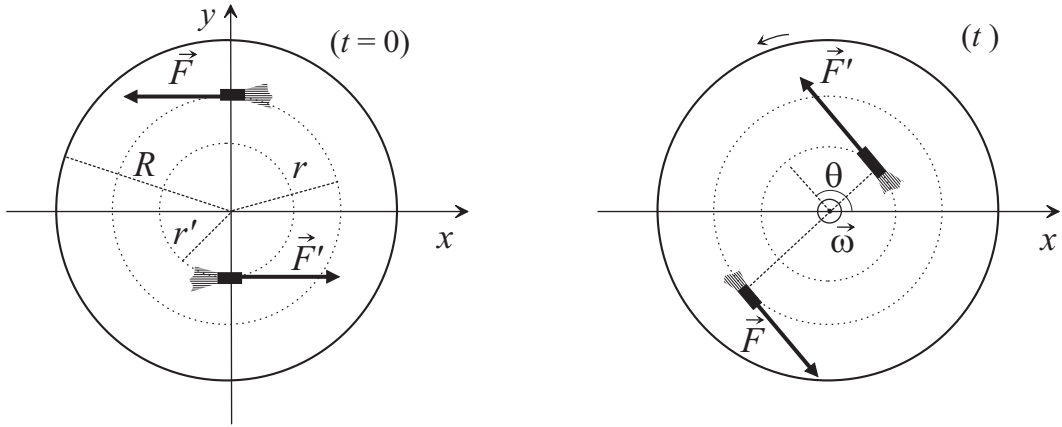}
\end{center}
\vspace*{-0.5cm}
\caption[]{\label{fig:fireworks} \small Fireworks wheel (with two cartridges of gunpowder), at $t=0$ and at a generic $t$.
Indicated are the forces on the wheel resulting from the combustion of the gunpowder.
}
\end{figure}

The next  system, illustrating the increase of rotational kinetic energy due to an internal source, belongs to a type of examples that is scarce in the literature.
In figure~\ref{fig:fireworks} we sketch  a simple fireworks wheel
with two cartridges filled with gunpowder.
When the system stars the operation, the rotational energy increase is obviously due to the chemical reactions in the gunpowder.

At a certain time, $t$, the wheel already rotated by an angle $\theta$ due to the binary of forces, $\vec F$ and $\vec F'$, resulting from the combustion of the gunpowder,  whose magnitudes are equal and assumed to be constant,
from $t=0$ until the exhaustion of the gunpowder. The cartridges containing gunpowder are glued to the wheel,
not necessarily equidistant from the center, as shown in  figure~\ref{fig:fireworks}. There is no translational motion, so that $\Delta K_{\rm cm}=0$  in (\ref{totale}). Moreover, the rotation around the axis is free
(no friction forces are assumed).

 The mechanical system is the rotating wheel (including the cartridges and excluding the products of the gunpowder chemical reaction, which anyway are almost massless). For magnitude constant  forces it is straightforward to apply equations (\ref{eq:in002a}) and (\ref{eq:in002b})
to the present example which, after integration, lead to
\beq
\left\{
\begin{array}{rl}
I \omega  &=  F (r+r')  t  \, , \vspace*{0.3cm}\\
\ \ {1\over 2} I \omega^2 &=  F (r+r')  \theta \, .
\end{array}
\right.
\label{86hr}
\eeq
Once more $\omega=2\theta/t$ and the constant angular acceleration, $\alpha=\omega/t$, is given by
\beq
\alpha = {F(r+r') \over I}\, .
\eeq

The energy equation (\ref{totale}) is obviously required to analyze this fireworks wheel.
For instance, the role played by the gunpowder has not been explicitly considered up to now but it enters, indirectly,  through $F$ and $t$ in equations~(\ref{86hr}):
the maximum $t$ is related to the amount of gunpowder in the cartridges
and $F$ is related to its `quality'.

Before applying equation (\ref{totale}) one should carefully define the thermodynamical system. To derive the previous mechanical equations, the mechanical system was essentially the wheel. The two forces $\vec F$ and $\vec F'$ are ``external" since they are, ultimately, produced by the gunpowder which was excluded from the system.
However, for the thermodynamical description, it is more interesting to include the gunpowder in the system. As already mentioned,
$\Delta K_{\rm cm} = 0$
and the first law of thermodynamics --- equation (\ref{totale}) --- for this process  is
\beq
\label{eq:fw003}
\Delta  U_{\rm R} +  \Delta U_T + \Delta U_\xi =W_\xi + Q_\xi\, .
\eeq
On the left-hand side we sum
the contributions to the internal energy variation due to the increase of rotational kinetic energy, to the temperature variation and to the
chemical reactions:
\beq
\left\{
\begin{array}{rl}
\Delta U_{\rm R} &= {1\over 2}I \omega^2 \\
 \Delta  U_T &= M c (T - T_0) \\
\Delta U_\xi &= n \Delta u_\xi\, ,
\end{array}
\right.
\label{ver45a}
\eeq
where $T_0$ is the initial temperature and $T$ the temperature at time $t$, $c$ is the average specific heat of the system of mass $M $, and $n$ is the number of moles of consumed
gunpowder.
 For simplicity, we firstly neglect the temperature variation in (\ref{ver45a}).
Let us now analyze the contributions to the right-hand side of (\ref{totale}).
A fraction of the energy liberated in the chemical reaction,  must be  spend as work, $W_\xi$, due to gas expansion against the external atmospheric pressure (a work reservoir, at constant pressure $P$),
and another fraction is spent as heat, $Q_\xi$, exchanged with the wheel surrounding  (a heat reservoir, at constant temperature $T_0$). Denoting by $\Delta v_\xi$ and $\Delta s_\xi$ the gunpowder chemical reaction specific volume and entropy variations, respectively,
the two terms on the right-hand side of equation (\ref{eq:fw003}) are explicitly given by
\beq
\left\{
\begin{array}{l}W_\xi = - n P   \Delta v_\xi \\
Q_\xi = n T_0   \Delta s_\xi\, .
\end{array}
\right.
\label{ver45b}
\eeq
 The molar internal energy, volume and entropy variations associated with the chemical reaction can be determined in some independent experiment. Else, they can be obtained in the framework of a model calculation. For instance, the molar internal energy variation $\Delta u_\xi$ is, in principle, computable from tabulated binding energies, knowing the number of broken and created bonds in the molecules that take part in the gunpowder combustion reaction.
One should note that there is, obviously, a contribution to the work in (\ref{totale}) from the forces $\vec F$ and $\vec F'$. However,  that contribution vanishes with the work of the corresponding
reaction forces exerted by the cartridges on the gunpowder. Since the gunpowder is now included in the system, there are two pairs of action-reaction forces, namely $(\vec F, -\vec F)$ and $(\vec F', -\vec F')$, and
the total work of each pair vanishes. In conclusion, when the system includes, altogether, the wheel and the gunpowder inside the cartridges, those action-reaction pairs are internal forces whose total work, in the present case, vanishes, since the displacement of the forces in each pair is the same.

Inserting  (\ref{ver45a}) and (\ref{ver45b}) in equation (\ref{eq:fw003}) one readily obtains
\beq
\label{eq:fw003b}
{1\over 2}I  \omega^2   = - n \Delta g_\xi\, ,
\eeq
where   $\Delta g_\xi = \Delta u_\xi + P\Delta v_\xi - T\Delta s_\xi$ is the  Gibbs molar free energy variation for the chemical reaction.
 It is the Gibbs energy variation that naturally appears in the energy balance equation because the process is assumed to take place  in contact with a heat and a work reservoirs. In a not so refined calculation it would be simply the internal energy variation and not the free Gibbs energy that would appear in the equations.
If the wheel increases its temperature during the process, there is an extra term, $Mc(T-T_0)>0$ on the left-hand side of equation   (\ref{eq:fw003b}), i.e. ${1\over 2}I  \omega^2   < - n \Delta g_\xi\, $
because part of the available energy ought to be used to increase the temperature of the wheel, so that the final kinetic energy would be correspondingly reduced.

But let's go back to equation (\ref{eq:fw003b}), which is the analog, for the rotations, to a similar expression for translations derived in the context of the study of an accelerating car \cite{guemez13b}.
 If initially the wheel is at rest, the angular velocity after the consumption of an amount $n$ of gunpowder is
\beq
\omega = \left({2 n \left|\Delta g_\xi\right|\over I}\right)^{1/2}\, .
\eeq
Using the second equation (\ref{86hr}) one also concludes that
\beq
   F (r + r') \theta = - n \Delta g_\xi\, .
\eeq

Finally, the amount of consumed gunpowder can be obtained as a function of time. For a given (constant) $F$  --- therefore, a constant angular acceleration, $\alpha$ ---, one finds
\beq
n= {F^2 (r+r')^2 \over 2 I \left| \Delta g_\xi \right| }\, t^2\, .
\eeq
For this idealized process,
the entropy of the universe does not increase, $\Delta S_{\rm U} = 0$, if the process is reversible. Actually,  the (free Gibbs) energy decrease resulting from the gunpowder reaction is stored in the system as rotational kinetic energy of the wheel. The system's capacity to
produce external work did not decrease. The variation of the Gibbs energy corresponds to the maximum ``useful work" that can be obtained. 
In the idealized situation, the equality holds in equation (\ref{eq:fw003b}) and the capacity of the system to produce useful work is totally kept. In the limit, that stored (organized) kinetic energy  can be used, in particular to increase by $\Delta G=-n \Delta g_\xi$ the Gibbs function of another chemical reaction elsewhere \cite{atkins10}. In this sense the process occurs without any entropy increase of the universe.

\section{Joule's apparatus}
\label{sec:joule}

\begin{figure}[htb]
\begin{center}
\hspace*{-0.5cm}
\includegraphics[width=7cm]{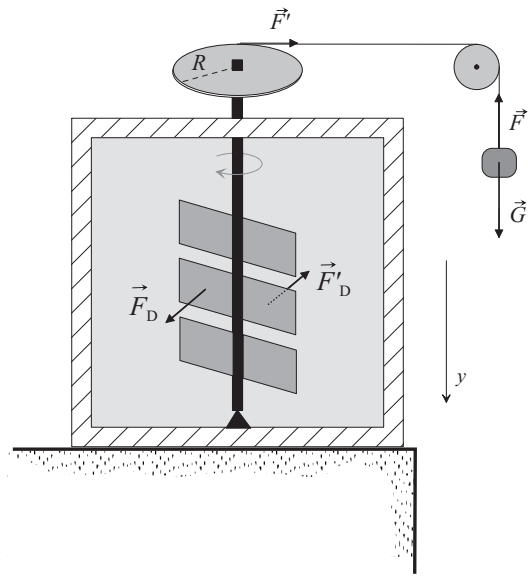}
\end{center}
\vspace*{-0.5cm}
\caption[]{\label{fig:pasdejoule} \small Joule's apparatus to measure  the mechanical equivalent of heat.
 On the right side we represent the descending block and the two forces applied to it: the tension, $\vec F$ and the weight $\vec G$; we also represent the force on the pulley transmitted by the rope, $\vec F'$ and the dragging forces exerted by the liquid on the paddles, $\vec F_{\rm D}$ and $\vec F'_{\rm D}$, which are along parallel lines but in opposite directions.
}
\end{figure}

The historic wheel paddle experiment was firstly performed by Joule  to measure the ``mechanical equivalent of heat"~\cite{zemansky97}. In a simplified version, it can be described as a wheel paddle that is immersed in a liquid (e.g. water) contained in a vessel with rigid adiabatic walls \cite{lemons09}. When it is placed in motion, the wheel paddle stirs the liquid. Figure \ref{fig:pasdejoule} represents Joule's apparatus:
a spindle attached to the paddles turns around due to the action of a rope, wrapped around its rim. The rope is connected, through a pulley, to a descending block of mass $m$.

The tensions $\vec F$ and $\vec F'$ in the (massless) rope are equal in magnitude  and this establishes a relation between the mechanical system on the left (spindle attached to paddles) and the mechanical system on the right (block). The block falls down subjected to its weight, $\vec G$, and to the upward tension whose magnitude is variable but such that $F\le mg$. The actual value of $F$ is also determined by the dragging forces ($\vec F_{\rm D}$ and $\vec F'_{\rm D}$) produced by the fluid on the paddles. The torque due to these forces, $\vec \Gamma_{\rm D}$, depends on the paddle's area, on the viscosity of the fluid and, surely, on the velocity of the rotating paddles. One expects that, after some time, a steady state for which $F=mg$ can eventually be achieved. Though $F$ is a variable force, its actual value is not needed for the forthcoming discussion.
The falling object communicates, via the tension in the rope, an external torque $FR$ to the paddle, where $R$ is represented in  figure \ref{fig:pasdejoule}.

Since $F$ is, at least for some time, a variable force, it is better to use the differential form of the equations to describe the process, as given in section 1. Let us first discuss the motion of the
block alone. The second equation in (\ref{eq-3}) leads, in the present case, to
\beq
(mg-F) \ \d y = {1\over 2} m \d v^2
\label{fga23a}
\eeq
Now let's consider the  other mechanical system: the  moving  wheel paddle. We assume that the rotation around the axis is frictionless. On the other hand, all vertical forces (weight, normal force) have zero resultant and they
do not contribute to the rotation since they are along the rotation axis. For the rotation of the wheel paddle there are two torques to be considered: the one produced by the tension $\vec F'$ and the torque produced by the fluid, in the opposite direction, resulting from the dragging forces represented in figure \ref{fig:pasdejoule}. Under the action of the two torques,  equation (\ref{eq:in002b}) takes the form

\beq
\left( FR  - \Gamma_{\rm D}\right) \d \theta= {1\over 2}I \d \omega^2\,
\label{fga23b}
\eeq
where $I$ is the moment of inertia for the rotation of the wheel paddle around the vertical axis  (with the minus sign on the left hand side, $\Gamma_{\rm D}>0$).
The connection between the translation of the block, on the right, and the rotation of the wheel paddle, on the left, is established by the equation $\d y = R \d \theta$ (or, equivalently, by $\d v = R \d \omega$).

Regarding equation (\ref{totale}) applied to the block, it does not add new information with respect to (\ref{fga23a}), but it does when it is applied to the wheel paddle and the liquid. For the thermodynamical discussion of the left part of figure~\ref{fig:pasdejoule} the system {\color{black} is now} the wheel paddle  {\em plus} the liquid inside the vessel. One should note that the forces exerted by the paddles on the fluid and by the fluid on the paddles
are equal and opposite (action-reaction pair) and these internal forces produce a vanishing work contribution to the right-hand side of (\ref{totale}). (The situation is similar to the one found in section~\ref{sec:fire} when we included the gunpowder in the system.)
Let us denote by $M$ the mass of the liquid in the vessel which is thermally isolated from the exterior. If we denote the specific heat of the fluid by $c$, the variation of the internal energy receives a contribution from the temperature change, $\d U_T= M\,c\, \d T$. The other contribution to the internal energy variation of the system is
due to the wheel paddle rotation, $\d U_{\rm R}= {1\over 2} I  \d \omega^2$. Since $\delta Q=0$ (vessel with adiabatic walls, so there is no external heat transfer) and the external work reduces to $F R \d \theta=F  \d y$, equation (\ref{totale}) takes the form (for the system under consideration --- paddles and liquid ---, $\d K_{\rm cm}=0$)
\beq
 {1\over 2}I \d \omega^2 +M \,c \, \d T = F\, \d y\, .
\label{fga23c}
\eeq
Eliminating $F$ by means of (\ref{fga23a}) one concludes that
\beq
M\, c \,\d T = m\,g\, \d y - {1\over 2} m \d v^2 - {1\over 2}I \d \omega^2
\eeq
or, eqivalently, $m\, g \, \d y=M\,c \, \d T+\d K$, where the second term here represents the variation of the total kinetic energy of the global system.
Is is worthwhile to point out that, before the steady state starts, the potential energy is also transformed into kinetic energy of the block and paddles.

When the steady state is achieved (constant $v$ and constant $\omega$ --- in the original Joule experiment $v \approx 0$ and $\omega \approx 0$), the previous equation can be integrated out, yielding, for a constant specific heat,
\beq
T= T_0 + {mg\over Mc} \Delta y
\label{sdfg}
\eeq
where $T_0$ is the temperature of the fluid when the steady state starts.
This equation expresses the direct transformation of `organized' gravitational potential energy
into  `disorganized' internal energy  with a consequent increase of the entropy of the universe. Of course, we are assuming that the paddles, as well as the vessel walls, do not change their temperature.

The increase of the entropy of the universe is only due to the entropy increase in the fluid. It can be easily calculated using an auxiliary quasi-static process (Clausius algorithm) \cite{leff12}
of heating up the fluid on the vessel:
\beq
\Delta S_{\rm U} = \int_{T_0}^T {Mc\d T \over T} = Mc \ln {T \over T_0} > 0
\eeq
or, if $T\sim T_0$, and using (\ref{sdfg}),
\beq
\Delta S_{\rm U} = {m\, g\, \Delta y \over T_0}\, .
\eeq
The process is clearly irreversible ($\Delta y>0$ for the descending block).

A final remark on equation (\ref{fga23b}). For the steady state it can equivalently be expressed, after integration, as
\beq
W _{\rm D} =mg\Delta y \ \ \ \ {\rm or} \ \ \ \ W _{\rm D}=M\, c\, \Delta T
\eeq
where $W_{\rm D}=\Gamma_{\rm D}\Delta \theta$ is the torque exerted on the paddles produced by the dragging forces times the angular displacement. This is also equal to the dissipative work of the dragging forces \cite{anacleto10}  that leads to the direct increase of the internal energy of the liquid, through a temperature increase. Therefore,
a `mechanical equivalent of heat' ({\em i.e.} dissipative work) was used to increase the temperature of a system. The result expressed by (\ref{sdfg}) was originally observed experimentally by Joule, and played an important role in the development of thermodynamics.

\section{Conclusions}
\label{sec:conclusions}

We presented and discussed various examples, all involving rotational motion, whose analysis requires the first law of thermodynamics in addition to the Newton's second law.
We started with a situation for which the mechanical treatment is sufficient. Such processes are reversible and do not lead to  an entropy increase of the universe. In general, this is not the case when
we are in the presence of dissipative forces, also analyzed in this paper. In particular we considered kinetic friction forces whose work is thermodynamically equivalent and considered as heat (therefore that work can set to zero  and included as heat). The role of friction forces is subtle and sometimes not adequately  addressed in textbooks \cite{bauman92}.
Examples with rotational kinetic energy increase due to internal sources were also considered here, in particular we discussed the motion of a fireworks wheel.

Finally, we discussed the Joule's wheel paddle apparatus, an historic experiment to measure the mechanical equivalent of
heat. Thermodynamics  textbooks mention the experiment but usually they do not provide the detailed physics explanation for the process. In particular we stressed the role of
the forces of the paddles on the liquid that produce a dissipative positive work.

In this concluding remarks one should also stress that, in the formulation of the first law of thermodynamics, the kinetic rotational energy of a system and the kinetic translational energy with respect to the center-of-mass were always considered as part of the internal energy of the system.

Summarizing, we illustrated that processes undergone by mechanical systems, usually also require a thermodynamical detailed analysis in order to be fully understood, a point of view adopted
 in the most modern physics textbooks  \cite{chabay11b}.
With the variety of examples, all involving rotations, we intended to show the  appropriateness of the first law of thermodynamics to complement the pure mechanical description of a system provided by the Newton's second law.

\end{document}